\documentclass[%
 reprint,
superscriptaddress,
 amsmath,amssymb,
 aps,prl
]{revtex4-1}

\usepackage{graphicx}
\usepackage{dcolumn}
\usepackage{bm}

\begin{document}

\preprint{APS/123-QED}

\title{Twisted Acoustics}

\author{Xue Jiang}
\author{Bin Liang}
\email{liangbin@nju.edu.cn}
\author{Jian-chun Cheng}
\email{jccheng@nju.edu.cn}
\affiliation{Collaborative Innovation Center of Advanced Microstructures and Key Laboratory of Modern Acoustics, MOE, Institute of Acoustics, Department of Physics, Nanjing University, Nanjing 210093, People’s Republic of China}
\author{Cheng-Wei Qiu}
\email{chengwei.qiu@nus.edu.sg}
\affiliation{ Department of Electrical and Computer Engineering, National University of Singapore, 4 Engineering Drive 3, Singapore 117583, Singapore}

\date{August 25, 2017}

\begin{abstract}
We use metasurfaces to enable acoustic orbital angular momentum (a-OAM) based multiplexing in real-time, postprocess-free and sensor-scanning-free fashions to improve the bandwidth of acoustic communication, with intrinsic compatibility and expandability to cooperate with other multiplexing technologies. The mechanism relied on encoding information onto twisted beams is numerically and experimentally demonstrated by realizing the real-time picture transfer, which differs from existing static data transfer by encoding data onto OAM states. Our study can boost the capacity of acoustic communication links and offer potential to revolutionize relevant fields.
\end{abstract}

\maketitle


Acoustic communication is pivotal in applications such as ocean exploration, where sound is the dominant information carrier due to the prominent loss of light in ocean \cite{ref:Stojanovic-1996,ref:Kilfoyle,ref:Kumar,ref:Singer}. Unlike optical communication \cite{ref:Gnauck} with the properties of high frequency and light speed, information transfer based on sound is subject to deficiencies of low frequency and velocity, limiting the development of more advanced acoustic communication \cite{ref:Stojanovic-1996,ref:Kilfoyle,ref:Kumar,ref:Singer}. Although remarkable progress has been made by introducing wavelength-division multiplexing (WDM), time-division multiplexing (TDM) and multilevel amplitude/phase modulation \cite{ref:Stojanovic-1994,ref:Ochi,ref:Freitag,ref:Li}, data-rate of acoustic communication is approaching its current limit, due to that sound, as a scalar wave, bears no polarization or spin, as opposed to its electromagnetic counterparts. It is stringent to exploit plausible multiplexing mechanisms to encode information in a scalar field with multiple states orthogonal and compatible to existing degree of freedoms (DOFs). Orbital angular momentum (OAM), with the infinite dimensionality of its Hilbert space and unbounded orthogonal states, is a promising candidate.

Many efforts have been made in optical OAM-based multiplexing \cite{ref:Gibson,ref:Shapiro}, such as spiral phase masks \cite{ref:Bozinovic,ref:wang}, Dammann gratings \cite{ref:zhang,ref:Lei}, q-plates \cite{ref:Nagali} or interferometers \cite{ref:Leach}. The precise control of sound would result in bulky device if we directly translate the optical mechanisms into acoustics. Previous works on a-OAM beams \cite{ref:Hefner,ref:Jiang} primarily exploit their mechanical effects such as for particle trapping and manipulation \cite{ref:Zhang-PRE,ref:Qiu-APL,ref:Hong,ref:Baresch}, only one work has attempted to utilize OAM in acoustic communication \cite{ref:shi}. However, the method in Ref.~\cite{ref:shi} relies on encoding data on OAM states and only employs this single dimension to realize a static data transfer, while in reality the fast and continuous information transmission should work in a real-time and dynamic manner. In addition, active sensors array is required to scan the spatial field and complex algorithms should be conducted for data decoding, which will not only impose extra loads to the existing communication link both in hardware and software, but also limit the transmission speed due to the time consumption in post data-processing. The active decoding method would also restrict the transmission accuracy since the bit error rate highly dependents on the transducers number in the sensor array\cite{ref:shi}.

\begin{figure*}
\includegraphics[scale=0.45]{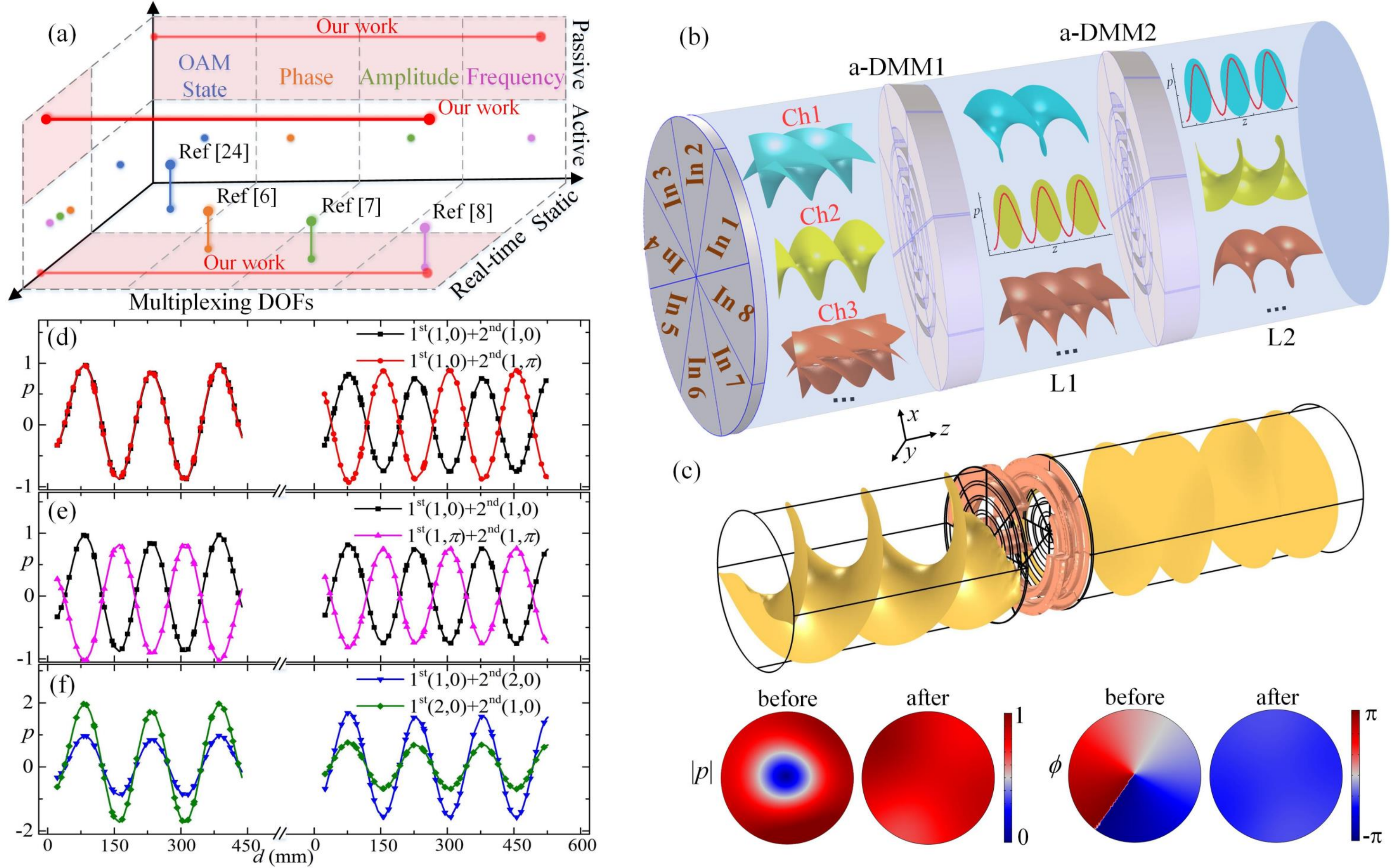}
\caption{\label{fig:1}(a) Comparisons among our work, the methods in Ref.~\cite{ref:shi} and other milestone references. (b) Schematic. For multiplexing, the entrance is divided into eight fan-like sections, with each accessing to an individual input. For de-multiplexing, twisted beams with null core are converted to updated a-OAM beams or plane wave (non-null core) according to their a-OAM values. (c) Equiphase surfaces of pressure $p$ for the $1^\mathrm{st}$ order a-OAM beam transmitting through an a-DMM in waveguide. (d)-(f) Normalized sapatial distributions of $p$ along the centerline in region L1 and L2 with different input multiplexing signals. a-OAM, acoustic orbital angular momentum; In: input; Ch, channel; a-DMM, acoustic de-multiplexing metasurface.}
\end{figure*}

We theoretically propose and experimentally validate the twisted acoustic beam with OAM for real-time information transfer in a passive, postprocess-free and sensor-scanning-free paradigm with metasurfaces \cite{ref:Sounas,ref:Ye,ref:Monticone,ref:Zheludev,ref:Jing-PRL,ref:Yong-PRL} to overcome those aforementioned issues. Rather than encoding data onto OAM states, here we use the a-OAM beams as the data carriers, exhibiting the instinct compatibility with pre-existing DOFs. We enter the null of the twisted beams and take full advantage of this trait, which is usually less significant. A subwavelength acoustic de-multiplexing metasurface (a-DMM) with a thickness of $0.5\lambda$ and radius of $0.53\lambda$ ($\lambda$ is sound wavelength, see Supplementary Material for details), as a passive and compact de-multiplexing component, is designed to directly and promptly decode data by a single transducer. Comparisons among the a-DMM-based information transfer in this work, the methods in Ref.~\cite{ref:shi} and other milestone references are shown in Fig.~\ref{fig:1}(a). The pressure transmittance is 91.5\% for one a-DMM and a nearly 100\% data transmission accuracy is achieved. Advantages of free of signal postprocess and field scanning, passive and compactness, high capacity and accuracy in real-time data transmission will be demonstrated both numerically and experimentally in what follows.

Schematic of the a-OAM multiplexing and de-multiplexing mechanism is illustrated in Fig.~1(b), where twisted beams with different topological charge $m$ propagate in an overlaid fashion as orthogonal channels. Note that OAM beam has a spiral phase dislocation $e^{im\theta}$ ($\theta$ is azimuthal angle) and null core, which is a crucial feature in our scheme. The multiplexing signal comprising of twisted beams with different $m$ can be expressed as $p(r,\theta,z,t)=\sum_mA_m(t)e^{i(m\theta+k_z z+\phi_m(t))}$, which propagates along $z$ direction in waveguide to avoid attenuation due to diffraction. Here $k_z$ is the wavenumber, the time-dependent amplitude $A_m(t)$ and phase $\phi_m(t)$ can be merged in the multilevel formats, suggesting the handy combination with pre-existing technologies (e.g.,~WDM and TDM) without introducing extra loads. We use the angular spectrum (AS) to yield the exact input in transmitting end (see Supplementary Material), similar with that for calculating the profile in acoustic hologram \cite{ref:melde}. The continuous azimuthal-dependent $p_\mathrm{in}$ is discretized by dividing the entrance into eight fan-like sections, individually accessing to eight inputs. As a result of the spatial multiplexing, an enhancement of spectral efficiency by a factor of $N$ can be expected, with $N$ being the total number of a-OAM beams.

For separating twisted beams with different $m$ and thereby decoding the data in each a-OAM channel, a mechanism based on acoustic resonance is proposed: $N$ layers of identical passive metasurfaces denoted as a-DMMs, capable of converting acoustic resonances to a-OAM, are impressed successively at the receiving terminal, with the output detected by a single transducer at the center after each layer. Each a-DMM has an OAM of -1, leading the order of all the transmitted a-OAM modes to be lowered by 1. Consequently, the order of a-OAM after the $n^\mathrm{th} (1 \le n\le N)$ layer is the linear superposition of the a-OAM value of the incident beams $m$ and the total a-OAM provided by these $n$ structures, reading as $m-n$. Considering the inherent ‘doughnut’-shaped intensity profile of a twisted beam, only by using $m$ layers of a-DMMs to convert the spiral phase pattern of a $m^\mathrm{th}$ order beam to a planar shape can we remove the azimuthal phase term and observe a non-zero intensity at the core. An example is illustrated in Fig.~1(c), which is the equiphase surface of the $1^\mathrm{st}$ OAM beam before and after an a-DMM in waveguide. On the contrary, beams with other a-OAM values still remain spiral phase and null core. Therefore, we can precisely detect the information encoded onto the $m^\mathrm{th}$ order beam exclusively after the $m^\mathrm{th}$ a-DMM by a single transducer where the updated a-OAM order is exactly zero.

We present a demonstration via merging the data in multilevel phase (differential binary phase shift keying, DBPSK) and amplitude (quadrature amplitude modulation, QAM) formats. The notation $m^\mathrm{th}(A_m,\phi_m)$ indicates the $m^\mathrm{th}$ a-OAM beam with the amplitude $A_m$ and phase $\phi_m$ hereafter. The input multiplexing signal is the superposition of the $1^\mathrm{st}$ and $2^\mathrm{nd}$ a-OAM beams carrying the objective data. Acoustic pressure $p$ along central axis of the waveguide in region L1 and L2 after de-multiplexed by the a-DMMs are illustrated in Figs.~1(d)-(f), where the multiplexing signals in the resonance frequency $f_0=2287$Hz of the a-DMMs carry different information, as function of the distance $d$ from output surfaces of the corresponding layers. The pressure transmittance is 91.5\% after one a-DMM, guaranteeing a high transmission efficiency. For better comparison, all the values are normalized by the maximum absolute pressure $|p|$ in $1^\mathrm{st}(1,0)+2^\mathrm{nd}(1,0)$ case. The objective data carried by the $1^\mathrm{st}$ and $2^\mathrm{nd}$ twisted beams are perfectly restored after de-multiplexed by the two a-DMMs in regions L1 and L2, respectively. For example, in Fig.~1(d) the received signals in L2 have a phase shift of $\pi$ which is exactly the phase difference of the $2^\mathrm{nd}$ a-OAM beams between the two input signals, and in Fig.~1(f) the amplitude in the $1^\mathrm{st}(2,0)+2^\mathrm{nd}(1,0)$ case is two times (half) of that in the $1^\mathrm{st}(1,0)+2^\mathrm{nd}(2,0)$ case in L1 (L2). These results verify that the orthogonality between the a-OAM modes effectively avoids mode coupling and the associated crosstalk in the spatially independent channels. Moreover, they substantially prove that the a-OAM states are also essentially orthogonal to the pre-existing dimensions of phase and amplitude. Meanwhile, due to the resonant nature and consequent high frequency selectivity of the metasurface \cite{ref:Jiang}, only the a-OAM beams in the resonant frequency $f_0$ of the particular a-DMM can be converted to plane wave and detected, which ensures the orthogonality with frequency as well. The frequency selectivity helps to simplify the terminal configuration when combined with the WDM technology in which complicated equipment is required to filter \cite{ref:Cimini}, increasing the speed and reducing the burden of the post data processing.

\begin{figure}
\includegraphics[scale=0.34]{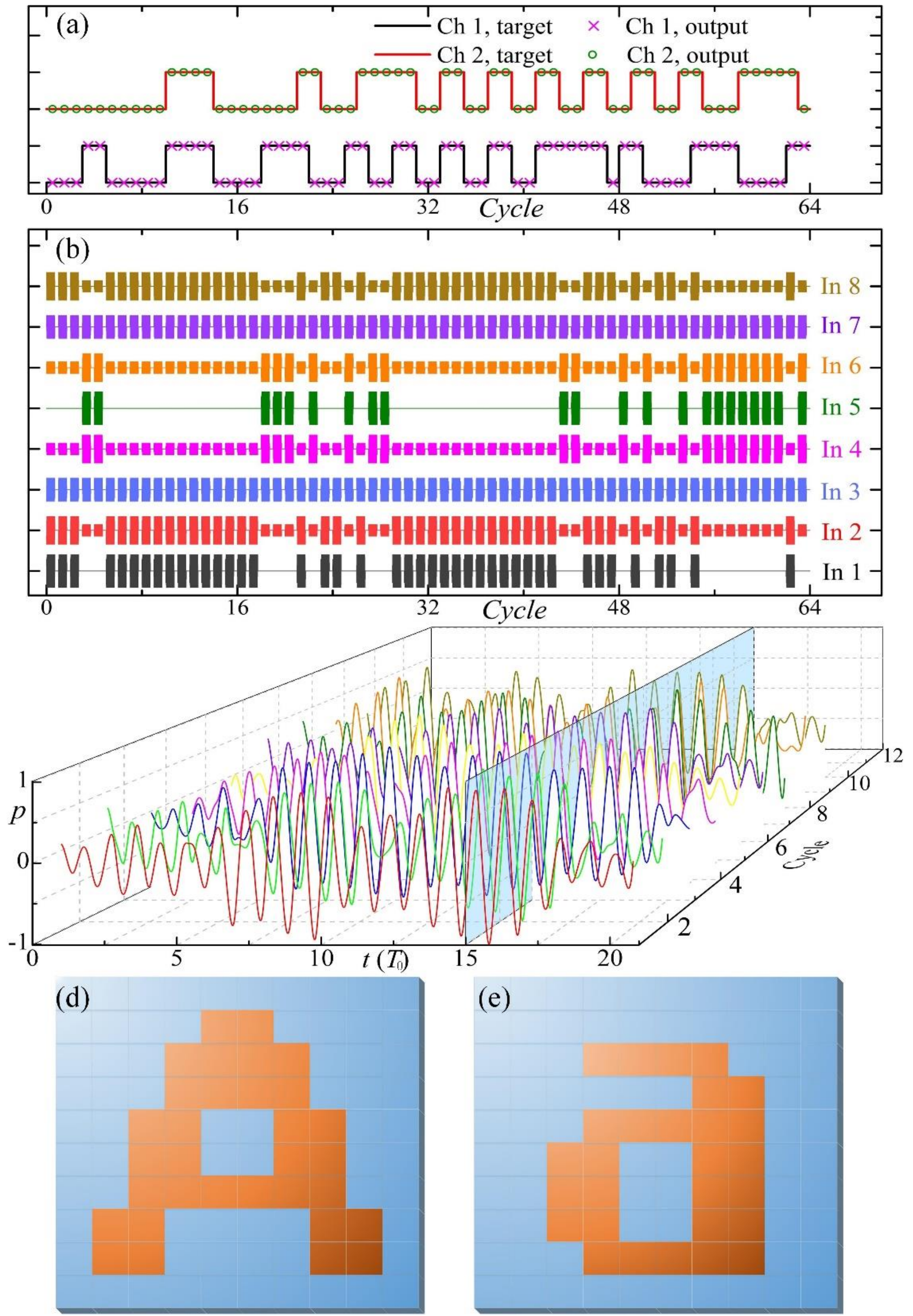}
\caption{\label{fig:2}Real-time data transfer. Multiplexing signal comprises of $1^\mathrm{st}$ and $2^\mathrm{nd}$ a-OAM beams serving as Ch1 and Ch2. (a) Comparison between the objective data and the output after de-multiplexed by the two a-DMMs. (b) Inputs in the eight sections [cf.~FIG.~1(b)] for generating the multiplexing signal, simultaneously enabling the two a-OAM channels. The signals are in pulse modulation, with the central frequency $f_0$ (period $T_0$), pulse period 20$T_0$ and duty ratio $0.7$. Each pulse cycle contains one-bit of data. (c) Part (the first 12 cycles) of the received signal as function of time in each pulse period in region L1. The blue plane is the reference surface for extracting information in each cycle of the pulse-modulated signal. (d)-(e) Images independently retrieved from the received data carried by a-OAM beams in Ch1 and Ch2.}
\end{figure}

We demonstrate the a-OAM multiplexing and de-multiplexing via real-time data transfer. As a visualized example, we use the $1^\mathrm{st}$ and $2^\mathrm{nd}$ a-OAM beams as two channels Ch1 and Ch2, and independently encode the pixels of two pictures, the images of letters ``A" and ``a", into the phase of a-OAM beams in DBPSK format through numerical simulations, where each pixel is encoded as a binary data (see Supplementary Material for simulation detail). For real-time communication, the multiplexing signal is in pulse modulation, with the central frequency $f_0$ (period $T_0$ and wavelength 15 cm), pulse period 20$T_0$ and duty ratio $0.7$, where each pulse cycle contains one-bit data. The eight inputs [cf.~Fig.~1(b)] for generating the multiplexing signal and simultaneously enabling the two channels are displayed in Fig.~2(b). In receiving terminal, two a-DMMs are cascaded to sort the $1^\mathrm{st}$ and $2^\mathrm{nd}$ a-OAM beams and two microphones (Mics) are placed to detect the signals in region L1 and L2. The received real-time signal in Ch1 by a Mic in L1 as function of time in each pulse period is partly (first 12 cycles) plotted in Fig.~2(c), where the blue plane is the reference surface for phase comparing and extracting the data in each cycle of the pulse-modulated signal. Data in other cycles and Ch2 is obtained in similar way. The decoded dataflows in Ch1 ($1^\mathrm{st}$ beam) and Ch2 ($2^\mathrm{nd}$ beam) are displayed in Fig.~2(a) in comparison with the objective. Two images are reconstructed with the received dataflows as shown in Figs.~2(d)-(e), which undistortedly reproduce the pictures of letters ``A" and ``a".

Experiments are conducted to verify the real-time communication based on a-OAM. Photographs of an a-DMM made of UV resin and experimental setup are shown in Figs.~3(a) and (b) (see Supplementary Material). Two a-DMMs, with the radius $0.53\lambda$ and thickness $0.5\lambda$, are sequentially inserted in cylindrical waveguide, and two Mics are placed centrally in L1 and L2. As a proof-of-concept experiment, we consider the independent transmission of two images each with $4\times4$ pixels encoded in the $1^\mathrm{st}$ and $2^\mathrm{nd}$ a-OAM beams. The experimentally received data extracted from the real-time transmitted signals (shown in Supplementary Material) in Ch1 and Ch2, comparing with the objective are displayed in Fig.~3(c). Images retrieved from the two dataflows are illustrated in Figs.~3(d) and (e) where perfect reconstructions are observed, demonstrating the experimental viability of the data transfer based on twisted beams.

\begin{figure}
\includegraphics[scale=0.34]{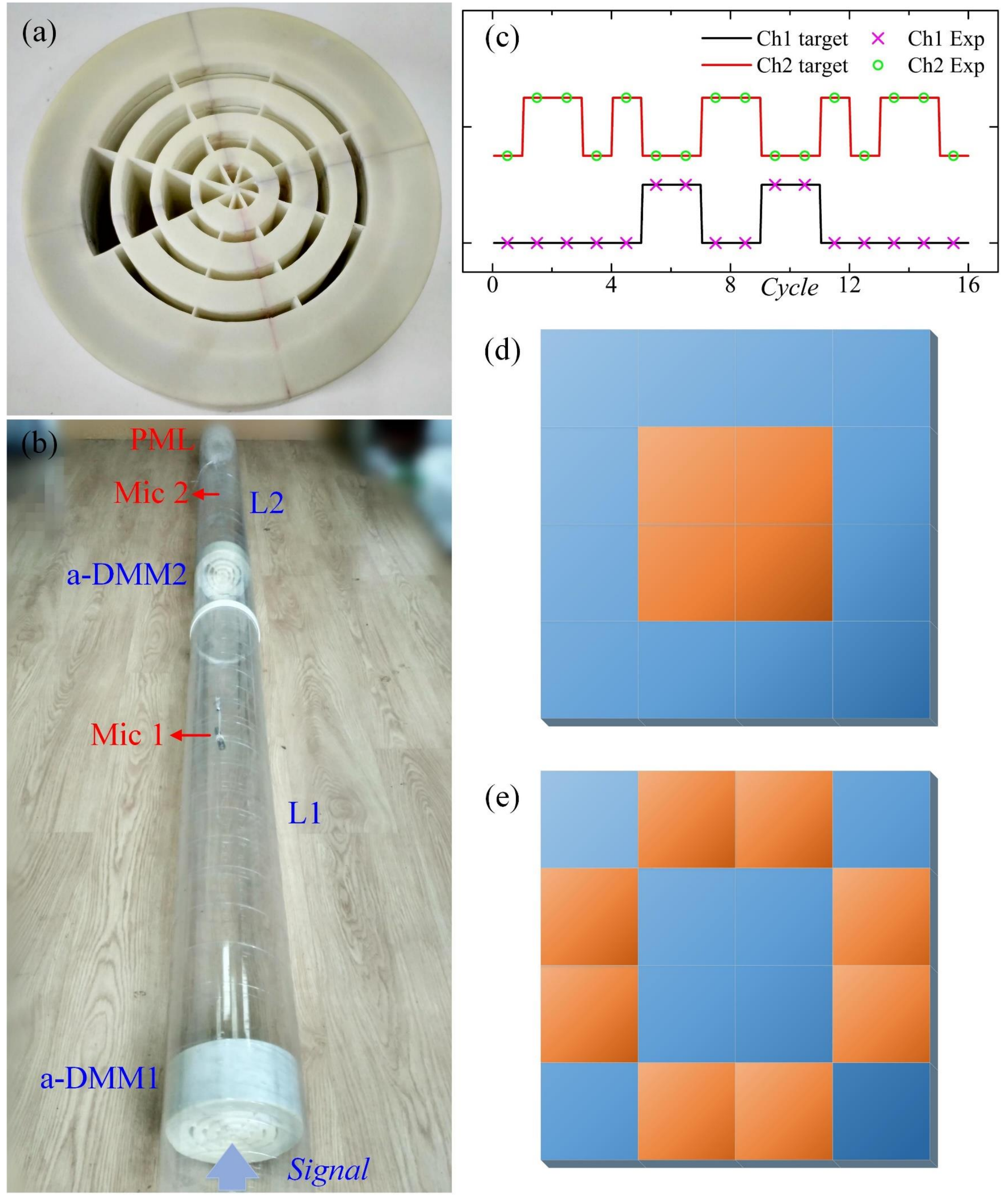}
\caption{\label{fig:3}Experiments. (a) Sample of the a-DMM. (b) Photo of the experimental setup. The multiplexing signal is synthesized and propagates in waveguide. Two a-DMMs are sequentially placed and two Mics are situated in the center of L1 and L2. (c) Experimentally received dataflows in Ch1 and Ch2 comparing with the objective data. (d)-(e) Retrieved images from the data received in Ch1 and Ch2. Mic, microphone; PML, perfect match layer.}
\end{figure}

Furthermore, we combine the a-OAM multiplexing with the multi-carrier modulation (MCM) technology to increase the transmission efficiency within the limited available bandwidth, which is crucial to transfer some urgent information and particularly beneficial in the varying fading conditions \cite{ref:Yee,ref:Ohtsuki}. The high-speed data stream is separated into several parallel flows of a relatively lower speed, encoded onto different a-OAM beams transmitting simultaneously, and then the decoded data are assembled accordingly. Here we demonstrate the MCM data transfer of the image of letters ``NJU", by encoding the image pixels alternately into the $1^\mathrm{st}$ and $2^\mathrm{nd}$ beams. The assembled data stream measured experimentally, in comparison with the simulation results and the objective are displayed in Fig.~4(a), where the inset partly shows the enlarged view of the comparison. Figure~4(b) shows the retrieved image with the received data, which is the exact reproduction of the original picture. The combination of MCM and twisted acoustics would facilitate the high-speed data transfer and improve the efficiency of post processing.

\begin{figure}
\includegraphics[scale=0.34]{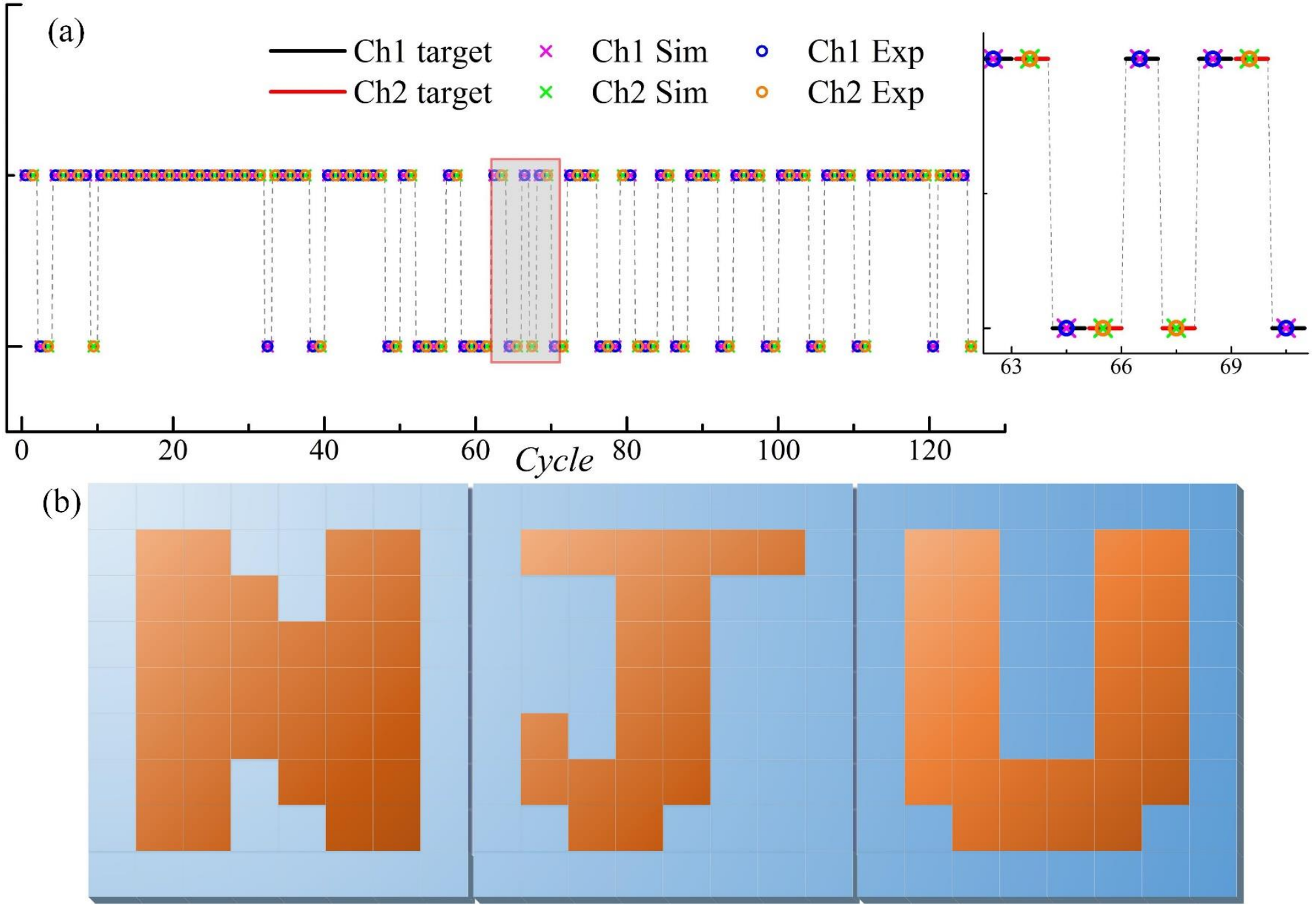}
\caption{\label{fig:4}Multi-Carrier Modulation (MCM). (a) Experimentally and numerically received high-speed dataflows, which are separately and parallelly encoded in Ch1 and Ch2 and assembled after independently de-multiplexed by the a-DMMs, in comparison with the objective. Inset, enlarged view of part of the data stream. (b) Image reconstructed from the assembled data stream both in simulations and experiments.}
\end{figure}

To conclude, we propose and experimentally demonstrate a simple and cost-effective scheme of twisted acoustics for real-time postprocess-free, sensor-scanning-free and high-capacity communication, where twisted beams of different OAM values serve as the spatially independent channels to carry information, with the compatibility with pre-existing multiplexing technologies. Subwavelength acoustic metasurfaces are designed as the passive and efficient de-multiplexing component for direct and prompt data decoding without sensors array and post-processing. We prove the effectiveness of the scheme via numerical simulations and experiments. It is noteworthy that the proposed scheme is universal since data transfer with this method is in principle not restricted by the number of OAM beams or transmitting distance, which can be extended to contain more a-OAM channels. Further improvements might lead to encode data by higher-order phase shift keying technology to achieve high capacity and spectral-efficiency acoustic communication.

In addition, the a-OAM based data transfer bears the security advantage of resisting to eavesdropping. Conventionally, the information would be covertly intercepted with additional receiver due to the atmospheric scattering, which require extra mathematical encryption. Our scheme offers the security enhancement as it is difficult to read the data without positioning the detector directly in the path of the intended receiver \cite{ref:Bozinovic}. In other words, the recovery is non-trivial. With the intrinsic orthogonality, high decoding efficiency and transmitting accuracy, increased integration density and the potential resistance to eavesdropping, twisted acoustics with OAM would take the acoustic communication to new heights, providing potential to improve the capacity and security of information transmission.
\begin{acknowledgments}
This work was supported by the National Key R\&D Program of China, (Grant No. 2017YFA0303700), National Natural Science Foundation of China (Grants No. 11634006 and No. 81127901) and a project funded by the Priority Academic Program Development of Jiangsu Higher Education Institutions. 
\end{acknowledgments}
\bibliography{reference}

\end{document}